\documentclass[conference]{IEEEtran}
\ifCLASSINFOpdf
  \usepackage[pdftex]{graphicx}
  \graphicspath{{./figs/}}
  \DeclareGraphicsExtensions{.pdf,.jpeg,.png}
\else
\fi
\newtheorem{conj}{Conjecture}

\begin{document}
%
\title{Do System Test Cases Grow Old?}

\author{

\IEEEauthorblockN{Robert Feldt}
\IEEEauthorblockA{School of Computing\\
Blekinge Institute of Technology\\
Karlskrona, Sweden\\
Email: robert.feldt@bth.se}

}

\maketitle

\begin{abstract}
Companies increasingly use either manual or automated system testing to ensure the quality of their software products. 
As a system evolves and is extended with new features the test suite also typically grows as new test cases are added.
To ensure software quality throughout this process the test suite is continously executed, often on a daily basis.
It seems likely that newly added tests would be more likely to fail than older tests but this has not been investigated in any detail on large-scale, industrial software systems. 
Also it is not clear which methods should be used to conduct such an analysis. 
This paper proposes three main concepts that can be used to investigate aging effects in the use and failure behavior of system test cases: test case activation curves, test case hazard curves, and test case half-life.
To evaluate these concepts and the type of analysis they enable we apply them on an industrial software system containing more than one million lines of code. The data sets comes from a total of 1,620 system test cases executed a total of more than half a million times over a time period of two and a half years.
For the investigated system we find that system \emph{test cases stay active as they age but really do grow old}; they go through an infant mortality phase with higher failure rates which then decline over time. The test case half-life is between 5 to 12 months for the two studied data sets.

\end{abstract}


%
\IEEEpeerreviewmaketitle

\section{Introduction}

\begin{figure}[t]
\centering
\includegraphics[width=2.5in]{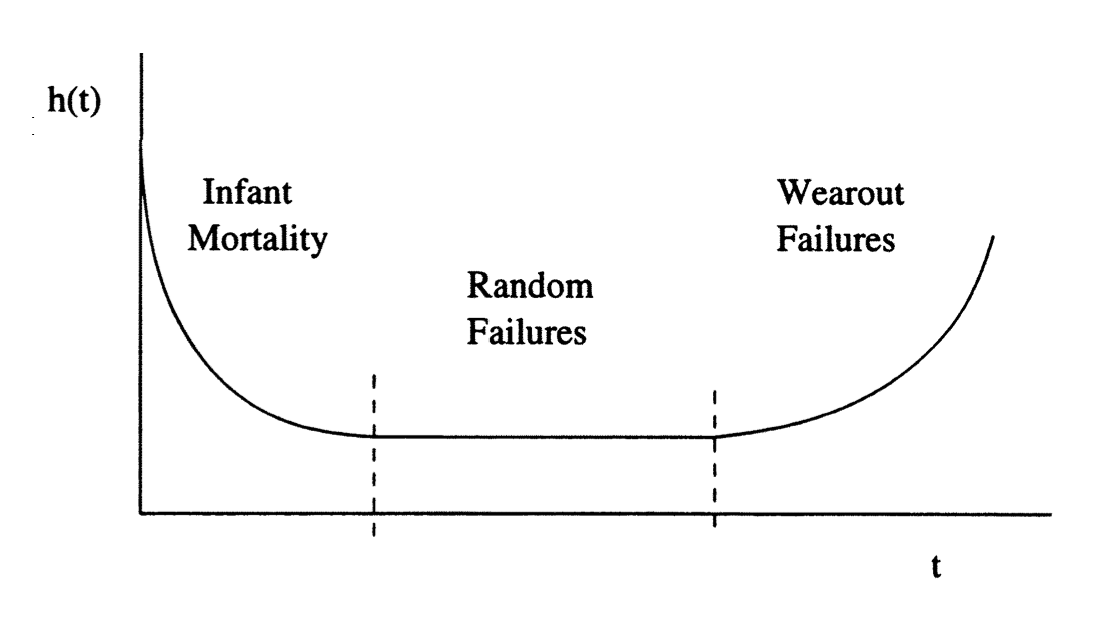}
\caption{Bathtub curve showing a commonly proposed failure rate over time for a physical component (from ~\cite{klutke2003critical})}
\label{fig_bathtub}
\end{figure}

The failure rate of hardware components, when graphed over time, is often claimed to take the shape of a `bathtub curve' with many initial failures, then a more stable period followed, finally, by increasing failure rates as the component grows old and wears out, see Figure~\ref{fig_bathtub}~\cite{klutke2003critical}. In this paper we want to find ways to study if any such `aging effects' can be found in the failure rate of software system test cases. Even though it seems counter-intuitive that a software system should `wear out' so that test cases would start failing more often at some point in time, an argument can be made that the test case failure rate should be higher for newly introduced tests since the system has not yet matured to handle the situations the tests cover.

In fact, the `bathtub curve' itself has been questioned even for hardware components~\cite{klutke2003critical}. Regardless of which shape(s) the actual failure curves for software test cases would have, if there are general patterns among them we could exploit this fact to better predict failures and prioritise testing activities. Possibly, we could even study the maturity of a test suite as a whole and provide decision support for when additional test cases needs to be added as the system evolves. As both the number of test cases and the amount of testing tends to increase over time, as a software system evolves, such knowledge could be broadly useful.

The prediction of defects and failures of software has been extensively studied over the years and with many different goals, e.g. for predicting fault-prone components~\cite{ostrand2005predicting}, predicting post-release failures via code metrics~\cite{nagappan2006mining}, and more generally to predict when quality and reliability is high enough so that testing can stop and the system be released~\cite{ledoux2003swrelmod}. There is also the related area of `software aging'~\cite{parnas1994software,grottke2008fundamentals} based on the empirical observation that software systems tend to be more likely to fail the longer they have been running, and the area of `online failure prediction' that bases system restarts on short-term predictions of the failure probabilities~\cite{salfner2010survey}. Many of these research areas rely on the more general, statistical methods of `survival analysis' to study the data at hand. However, there is a lack of studies that empirically study these aging effects on the test cases used during the development of a software system and it is not clear which statistical methods would support such studies. In particular, there is a lack of studies on large, industrial software systems.

In this paper we propose that three issues are fundamental in discussing test case aging: test case age, activity and effectiveness. We present methods to analyse and graph these concepts and help identify patterns. The proposed approach can be applied when the outcome of multiple executions of the same test cases have been collected over time. To evaluate the methods and start building an empirical basis for test case aging we evaluate our approach on a large-scale, industrial case system.

In Section~\ref{sec:rw} we give a more detailed background and describe related work. Section~\ref{sec:tc_aging} then presents the fundamental concepts for test case aging and the associated analysis methods. Section~\ref{sec:tc_case_study} describes the industrial case study and then applies the proposed methods on the extracted testing data. Finally, Section~\ref{sec:discussion} discusses validity threats while Section~\ref{sec:conclusions} concludes.

\section{Related Work}
\label{sec:rw}

While there is a lack of studies on aging effects in relation to software tests there is an abundant literature on the aging of software systems themselves (software aging) and how to model their quality and reliability over time (software reliability modeling and failure prediction). Below we provide a brief background on related work but point to surveys for more details.

The analysis of time series and event data are large sub-fields of mathematics and statistics with several relevant and recently published text books available~\cite{lutkepohl2005new,madsen2008time,miller2011survival}. Even though it is the basis for many of the engineering areas considered below we do not review their methematical foundations in any detail.

\subsection{Software defect prediction}

Software defect prediction models have been shown to be effective in several empirical studies. Ostrand et al~\cite{ostrand2005predicting} successfully predicted which files in a new software release were most likely to contain the largest number of faults. Their model was a negative binomial regression model based on the source code of the file as well as the failure and modification histories.

Nagappan et al ~\cite{nagappan2006mining} showed that when extensive data was available they could use sets of code complexity metrics to predict post-release software failures. However, Zimmermann et al~\cite{zimmermann2009cross} showed that defect prediction can be a challenge for new projects since prediction models could not easily be moved between projects. Recent benchmark studies in defect prediction that also give a good overview of the field can be found in \cite{Lessmann2008Benchmarking,dambros2012evaluating}.

\subsection{Reliability modeling and failure prediction}


Much research over the years has gone into predicting the failures, or related metrics such as quality/stability/reliability, of a software system based on data from its development, design or execution. When the prediction is about the long-term trends, e.g. whether the quality of the software as a whole increases or is high enough, the area is typically named \emph{reliability modeling}~\cite{salfner2010survey} but could be called offline failure prediction. When it is about the short-term prediction of failures based on monitoring the system state it is called \emph{online failure prediction}~\cite{salfner2010survey}.

Salfner et al~\cite{salfner2010survey} present a taxonomy of online failure prediction approaches based on what is the source of information on which the prediction is based: \emph{testing} to identify faults, \emph{auditing} (for undetected errors), symptom (side effects of errors) \emph{monitoring}, \emph{reporting} of detected errors, and \emph{failure tracking}. Their survey then focus on the latter four types but disregards failure prediction based on testing since it is typically not done online, during the execution of the software. However, online prediction algorithms share many similarities with offline ones and applies mathematical and statistical techniques such as time series analysis, system modeling, anomaly detection, and, recently and more generally, data mining and machine learning techniques to collected data related to the system state or failures. It also typically predicts the time between failures or time to the next failure for a sub-component or the system as a whole. The same is typically true in reliability modeling~\cite{ledoux2003swrelmod} (offline failure prediction) even if the reliability modeling area as a whole have also considered failure frequency models, i.e. failures per time period~\cite{farr1996software}.

A major difference between different software reliability modeling approches is whether they are black-box (not considering the inner structure of the system) or white-box (taking the design and architecture of the software into account)~\cite{ledoux2003swrelmod}. While there are methods for white-box modeling they are often considerably more complex and, in~\cite{ledoux2003swrelmod}, Ledoux warns that the possibly more predictive models they can generate may not be warranted by the increased modeling costs. He adds that developing better white-box modeling approaches would require a cooperation of the statistical modeling and software engineering communities which has historically not been common.

Go\v{s}eva-Popstojanova et al~\cite{goseva2005large} present an empirical study that evaluated white-box, architecture-based, software reliability models on an open-source, 350,000 lines of code application implemented in C. Results showed that several models were accurate when failures was due to faults in single components but all models failed for failures that was due to interactions between multiple components. 
In a related study on an industrial control system with more than 3 million lines of code, Kozioloek et al~\cite{koziolek2010large} conclude that architecture-based software reliability models are `still hard to apply' and that the tested method were not cost-effective since its benefits could not be quantified.

Pham~\cite{pham2006system} discusses the differences between hardware and software reliability engineering and concepts and notes that the `burn-in state (of hardware reliability) is similar to the software debugging state'.

An important statistical basis for reliability engineering is survival analysis~\cite{miller2011survival}. It is a branch of statistics that deal with the analysis of the time to some event, such as failures in engineered systems or deaths in biological organisms. Not only fatal or total failures can be studied, also partial failures or failures to one and the same subject or component can be modeled. Even though survival analysis typically assumes that one subject only fails once, which would make it less useful for software testing where test cases are typically used repeatedly, `recurrent event survival analysis' handles cases where the event occurs more than once per subject~\cite{kleinbaum2012recurrent}.

\subsection{Software aging}

Software aging is the term used to describe that the longer a software system runs the higher its failure rate is typically observed to be~\cite{parnas1994software,grottke2008fundamentals}. For hardware systems these types of effects are well-known and captured in concepts such as the `bathtub curve', i.e. that physical components goes through distinct phases during their lifetime: `infant mortality' where there are more failures early in the lifetime, `random failures' where the failure rate is constant, and `wear-out failures' when the component gets old and wears out. Even though the origin of this curve is unclear and it has been criticized~\cite{klutke2003critical} it is commonly introduced and built on in reliability engineering textbooks and research. Recent results in software aging typically focus on detecting trends in software system resource usage and activity parameters so that rejuvenation actions can be scheduled, see for example~\cite{grottke2006analysis}. Typical parameters are memory use and response times.



\section{Software Test Case Aging}
\label{sec:tc_aging}

We propose that the term `test case aging' is used to describe the set of concepts and analysis methods that study how (individual and) collections of software test cases evolve over time. The evolution over time can relate both to the number of, content of as well as failure finding ability of test cases. A large number of concepts and analysis methods could be useful that all, in some way, address the question of how (collections of) test cases evolve as they grow old. 

A basic question is how old test cases become, i.e. their age when they `die', i.e. are retired from a test suite\footnote{We use `test suite' to refer to any logically related set of test cases, e.g. all test cases for a whole system or for some sub-system.}. Another question concerns how active they are as they grow older. And maybe the most natural and pressing question is how effective they continue to be; do they still lead to failures and thus can be effectively used to identify faults. Below we will introduce three main types of analysis that correponds to these fundamental issues:

\begin{itemize}
  \item{Age - How old do test cases become?}
  \item{Activity - Are old test cases active, i.e. used for testing?}
  \item{Effectiveness - Are old test cases effective, i.e. do they fail at the same rate\footnote{An alternative view of test case effectiveness is if they help create a belief in the system as having a high quality, i.e. are used for validation rather than for verification. We do not consider this further here but leave it as future work. However, we think it would be hard to analyze objectively since the concept of `value for validation' which is likely to underlie any such endeavor would have to be subjectively measured.}?}
\end{itemize}

Other questions and issues that are important when considering test case aging effects are likely to surface but we consider these fundamental and focus on them in the following.

When analysing aging effects we will typically graph, model or statistically test an attribute (or statistic of an attribute) as it changes over time. Thus, we can identify two main axes of variation in any test case aging analysis in addition to which attribute is being studied and how: (a) unit of analysis (individual test case or set of test cases) and (b) time scale (wall-clock time or age-relative time). In this paper, we primarily focus on summary statistics for \emph{sets of test cases} and use an \emph{age-relative time scale}. This means that we align the data per test case based on the amount of time since each test case was created (it's age) and then calculate summary statistics for the set as a whole as a function of that age. However, our delimitation to one primary analysis type is for brevity's sake; future test case aging analysis is likely to need to look at other as well as use a combination of analysis types.


For the concepts and methods we introduce to be as generally applicable as possible we assume that only a very restricted, almost minimal, set of information about the studied testing and software development is available. In particular, we require no access to any of the development artefacts themselves (such as requirements, designs, source code, or, even, test cases). Rather, we only request meta-information about the testing such as when the test cases have been executed and what the outcome was. In our discussions with several industry partners this is an essential requirement. To get access to the actual development artefacts is much more sensitive and requires elaborate discussions about and then signing of non-disclosure agreements; for meta-data this is less of a problem.

Thus, before we detail the concepts and analysis methods, we present the restricted information model, called the `Creation-Execution-Outcome (CEO) information requirements model for test case aging analysis', that all of them require. The CEO model requires two main types of information which is outlined and exemplified in Tables~\ref{tab:first_test} and \ref{tab:test_execs}. The former contains information about the time when a test case was \emph{created} while the latter collects information about as many \emph{test executions, and their outcomes,} as possible. We assume the same test case names, denoted $c_i$, are used to refer to test cases for both types of information. For example, in Table\ref{tab:first_test}, there is a total of $N$ test cases having been added with test case $c_1$ named "A-1" and $c_2$ named "B-3". They were both created in August of 2007. Table \ref{tab:test_execs} then shows all the $M$ times that those same $N$ test cases from Table~\ref{tab:first_test} has been executed in the time interval from $t_1$ to $t_M$ as well as all the outcomes of running these tests. We assume the simplest possible fault model in which the outcome is either FAIL (also denoted F in the following) or PASS (P); all other outcomes are considered invalid and either filtered from the input data or mapped to one of these two valid outcomes\footnote{In our experience, different organizations use different levels of granularity when judging the types of outcomes. It is common to make a distinction between different reasons for why a test case was not executed or why it failed. For example, it might be due to a problem in the testing environment rather than in the system being tested. In the industrial case reported later we simply filtered out any such outcomes; however, very few such situations occured.}. 

\begin{table}[!t]
\renewcommand{\arraystretch}{1.3}
\caption{Creation (C) time for each test case}
\label{tab:first_test}
\centering
\begin{tabular}{c||c}
\hline
\bfseries Test case name & \bfseries Creation time\\
\hline\hline
$c_1$ = "A-1" & $t_0^{c_1}$ = 2007-08-25 10:00:10\\
$c_2$ = "B-3" & $t_0^{c_2}$ = 2007-08-27 15:23:37\\
... & ...\\
$c_i$ = "C-243" & $t_0^{c_i}$ = 2008-11-01 03:05:01\\
... & ...\\
$c_N$ = "D-42" & $t_0^{c_N}$ = 2011-07-06 04:57:47\\
\hline
\end{tabular}
\end{table}

\begin{table*}[!t]
\renewcommand{\arraystretch}{1.3}
\caption{Data on test executions (E) and outcomes (O)}
\label{tab:test_execs}
\centering
\begin{tabular}{c||c||c||c}
\hline
\bfseries Test case name & \bfseries Execution time & \bfseries Outcome & \bfseries Session start time \\
\hline\hline
$c_1$ & $t_1$ = 2007-08-25 20:00:10 & $o_{t_1}$ = PASS & $t_{s_1}$ = 2007-08-25 20:00:00 \\
$c_1$ & $t_2$ = 2007-08-26 20:00:10 & $o_{t_2}$ = FAIL & $t_{s_2}$ = 2007-08-26 20:00:00 \\
$c_1$ & $t_3$ = 2007-08-27 20:00:10 & $o_{t_3}$ = PASS & $t_{s_3}$ = 2007-08-27 20:00:00 \\
$c_2$ & $t_4$ = 2007-08-27 20:00:23 & $o_{t_4}$ = FAIL & $t_{s_3}$ = 2007-08-27 20:00:00 \\
...   & ...                         & ...              & ...                             \\
$c_i$ & $t_j$ = 2008-11-01 03:05:01 & $o_{t_j}$ = PASS & $t_{s_k}$ = 2009-11-07 20:00:00 \\
...   & ...                         & ...              & ...                             \\
$c_N$ & $t_M$ = 2012-01-06 04:57:47 & $o_{t_M}$ = PASS & $t_{s_Q}$ = 2012-05-28 20:00:00 \\
\hline
\end{tabular}
\end{table*}

For the test execution data we optionally also support information about the test session as a label on all the executions. In Table~\ref{tab:test_execs} the session information is shown in the rightmost column. A test session is identified by the point in time in which a set of test cases where evaluated on one and the same version of the system under test (SUT). The session information can be used to pre-filter the data. For example, in our industrial case there were sometimes problems with the building of the SUT or in the test execution framework and this was indicated by all or a large majority of test cases all failing for that same session. By having information about the session we can pre-filter so that such outliers do not negatively affect the further analysis. In the example shown in the table there are a total of $Q$ sessions.

Note that two main and different time intervals are implicit in this data. The test case creation interval, $T_C$, goes from $t_0^{c_1}$ to $t_0^{c_N}$ while the test execution interval, $T_E$, goes from $t_1$ to $t_M$. The two intervals might be the same but we make no such assumption; they might only partially overlap or not overlap at all. For example, in industry it is common that many test cases had been created before complete logging of test executions was in place or there may be missing information due to other causes.



\subsection{Test Case Age}

Naturally, we define the age of a test case as the time since its creation. However, test cases might become obsolete so to talk about a terminal age of a test case requires that we can determine a point of death, denoted $t_{\omega}^{c_i}$. We consider a test case to have died if it is no longer actively used for testing. Technically, we have that the age of a testcase $c_i$ at wall-clock time $t$, denoted $\alpha_{c_i}(t)$, is defined as $\alpha_{c_i}(t) = max(0, min(t, t_{\omega}^{c_i}) - t_0^{c_i}))$.

In some organisations the test management or logging databases contains information about which test cases have died. However, this information might not be available and we want to have objective methods to determine such a point of death, also in our restricted CEO model, in order for our analyses to be as widely useful as possible. To address this we propose a simple `test case death determination' scheme called last-execution-with-grace:



\begin{enumerate}

\item{
  Determine the last execution time, $t_{\omega}^{c_i}$, of the test case $c_i$, i.e. the largest time $t_j^{c_i}$ of the times when $c_i$ was executed.
}

\item{
  If $t_M - t_j^{c_i} \leq \gamma$, where $\gamma$ is a pre-selected grace period, we consider the test case alive, if $t_M - t_j^{c_i} > \gamma$ the test case is considered to have died. 
}

\end{enumerate}

We use $(t_0, t_{\omega})^{c_i}$ to denote the life span of the test case $c_i$. 
Given the life spans of all the test cases we can now, for example, create the growth curve for a given test data set by plotting for each time $t$ the number of test cases $c_i$ for which $t_0^{c_i} < t$ or $t < t_{\omega}^{c_i}$. However, this is an analysis done on a wall-clock time scale and in the following we do not consider it further. The definition of test case age is central for all analyses done on an age-relative time scale: for them we tally up information about individual test cases based on the times when they have identical age, rather than on identical wall-clock time.

The test case life spans also gives us a simple way to plot a histogram or densities of all the ages of test cases in the test suite since the age of a test case $c_i$, without reference to a specific time at which the age is determined, is simply $\alpha_{c_i}=t_{\omega}^{c_i} - t_0^{c_i}$. 

The introduction of a grace period is needed in the CEO model but also parameterizes subsequent analysis; a different grace period might give different results. In our experience, the grace period has only a small effect in practice though and it can be selected based on knowledge of the organisation or the frequency of testing. In the following, we will use a default grace period of 90 days, roughly three months, which corresponds to about 10\% of the length of the studied time interval $T_E$ in the industrial case.

\subsection{Test Case Activation Curves}

We propose to graph the probability that test cases are executed as a function of time to indicate the activity level of test cases. For the time scale we use the age of a test case, i.e. age-relative time, rather than wall-clock time. This way we can investigate if there are any patterns in how the execution probability evolves as test cases age, rather than as the system or test suite as a whole ages, as would be the case if we would have used wall-clock time.

Technically, we define the (empirical) execution probability (also called the activation rate) at (age) time $t$, denoted $p_{\epsilon}(t)$, as


\[
  p_{\epsilon}(t) = \frac{\# \{ c_i : t = t_j^{c_i} - t_0^{c_i} \wedge (o_{t_j^{c_i}} = F \vee o_{t_j^{c_i}} = P) \}}{\# \{ c_i : t \leq t_{\omega}^{c_i} - t_0^{c_i} \}}
\]

i.e. as the proportion of test cases that execute at (age) time $t$ over the number of test cases that have an age that is at least $t$. The $\#$ function returns the size of a set of objects. By graphing this curve as a function of age we can study if there are any patterns in how test cases are active as they grow older. We call such curves (empirical) test case activation curves (TACs). Note that a very similar curve can also be created based on wall-clock time if we consider which test cases are active at a point in time compared to how many test cases are alive in the test suite at that same point in time. However, using the transformed and aligned time scale (of age-relative time) allows us to compare test cases as they age and talk about the average age-based behavior.



\subsection{Test Case Hazard Curves}

A hazard function is typically defined as a measure of the tendency to fail, i.e. it plots the failure probability as a function of time. The greater the value of the hazard function, the greater the probability of a failure. The hazard function is sometimes also called the instantenous failure rate.

In reliability engineering, hazard functions are used to model the failure rate of system components. The failure rate is typically defined as the probability, $f(t)$, that a component would fail at time $t$, given that it has not failed at any time before $t$.

While a software test case is seldom a component that is released as part of the software system it would be useful if we would know the hazard function also of a test case. In particular, it would be important to study if such a test case hazard curve has a pattern over time. For example, it can be argued that test cases should be more likely to fail when they are young since the software developers have had less time to `harden' the system against the kind of challenge that the test case poses. For clarity lets formulate this as a conjecture\footnote{The term `mortality' is somewhat misleading for test cases since a test case does not `die' even if it fails, rather a test case that fails would be more likely to be run again, i.e. to be kept `alive'. Thus we could call the conjecture `test case infant clumsiness' to more accurately convey the idea (test cases are more likely to stumble when young). However, we have opted to use the term `mortality' since it is commonly used in the reliability engineering literature and is a well-known concept.}
:

\begin{conj}[Test Case Infant Mortality]
A test case is more likely to fail when it is young and, consequently, less likely to fail as it becomes older.
\end{conj}

If we could establish hazard curves empirically we could formulate this conjecture as concrete hypotheses and test them. However, it is not even clear how test case hazard curves could be established in practice. In system reliability engineering hazard functions can be established only from a population of components. In the words of Pham~\cite{pham2006system}: `The importance of the hazard function is that it indicates the change in the failure rate over the life of a population of components by plotting their hazard functions on a single axis.' The components are typically also considered to be homogeneous, i.e. trying to accomplish the same thing and being produced by the same (physical) process. None of these conditions apply for system test cases that are typically created to test (and thus accomplish) fundamentally different properties (and goals) and being produced by different processes. Still, as an overall statistical characterisation of how a set of test cases behave, producing such curves can help create insight. They key transformation is to transform wall clock time to a relative time scale based on the age of a test case; by aligning the individual curves for each test case we can then calculate statistics and produce summary graphs, for the test cases considered as a single group, on the transformed time scale.

Thus, technically, we define the (empirical) failure rate (probability) at age $t$, denoted $p_{\zeta}(t)$, as


\[
  p_{\zeta}(t) = \frac{\# \{ c_i : t = t_j^{c_i} - t_0^{c_i} \wedge o_{t_j^{c_i}} = F \}}{\# \{ c_i : t = t_j^{c_i} - t_0^{c_i} \wedge (o_{t_j^{c_i}} = F \vee o_{t_j^{c_i}} = P) \}}
\]

i.e. as the proportion of test cases that fail at (age) time $t$ over the number of test cases that execute at (age) time $t$. By graphing this curve as a function of age we can study if there are any overall patterns in how test cases fail as they grow older. We call such curves (empirical) hazard curves (HACs) or (empirical) failure rate curves. In the same way as for the activation curves, failure probability curves can also be created based on wall-clock time.

To characterise the overall shape of empirical hazard curves we propose to use simple linear regression model fitting~\cite{myers1990classical}. Linear or quadratic models as a function of the test case age should be the primary targets since they are easier to interpret and understand, however more complex model familys can be explored if greater predictive power is needed or theoretically motivated.

Hazard curves such as the bath-tub curve often go through periods of decrease that can be modeled with exponential functions, typically termed exponential decay. A concept that is often used to characterise such decay in chemistry and physics is \emph{half-time}. It is the `amount of time required for a quantity to fall to half of its value as measured at the beginning of the time period'~\cite{wiki2013halflife}, i.e. the smallest time $t_i$ at which $f(t_i) \leq f(t_0)/2$ where $f(t_0)$ is the value of a time-dependent function at the initial (first) time instance $t_0$. Since it is such a well-known concept it has been used also to characterise also decay functions that are not modeled as exponential decay. However, such use makes less sense if the decay is not monotonously decreasing since there can thus be multiple points in time that could be arguably named as the half-life points. In the following, we will use the term `test case half-life' for the half-life of empirical test case hazard curves that are, or at least can be reasonably be modeled as, monotonously decreasing.


\section{Industrial Case Study}
\label{sec:tc_case_study}

To evaluate the analysis methods presented in this paper and to start building a knowledge base on test case aging, we conducted a set of analyses on the testing data from a large, industrial software system developed by Ericsson AB in Sweden. The system has been in development for more than 6 years and is developed with a process heavily influenced by agile methodologies in a scrum-of-scrums configuration of multiple, cross-functional teams~\cite{sutherland2007distributed}. In 2012 its source code was made up of more than 1 million lines of (mainly) Java code. The development teams use continuous integration with automated testing at several test levels, from unit testing, via integration testing and up to system testing levels. The market that the system targets has traditionally been focused on system testing, in particular, black-box testing of externally observable features and functions. But in the last couple of years they have increasingly also used automated testing at lower levels. The system tests that we have investigated are partly written by an independent testing team but are also written by developers themselves. System tests are written primarily in TTCN-3~\cite{schieferdecker2003uml} and automatically executed both nightly and in different subsets during code commits and testing checkpoints during the day.

\begin{table}[!t]
\renewcommand{\arraystretch}{1.3}
\caption{The two data sets used in the empirical study}
\label{tab:datasets}
\centering
\begin{tabular}{l||c||c||c}

\hline
\bfseries Attribute & \bfseries Main data set & \bfseries Validation data set & \bfseries TOTAL\\

\hline\hline
System branch   & A       & B         & 2 branches       \\
Test groups     & C, D, E & F, G, H   & 6 groups       \\
Num. test cases & 1,040    & 580       & 1,620   \\
Num. executions & 421,240 & 94,701    & 515,941\\
Earliest execution, $t_1$ & 2009-10-29 & 2011-10-20 & \\
Latest execution, $t_M$ & 2012-05-28 & 2012-05-28 &   \\
Earliest test creation, $t_0^{c_1}$ & 2008-06-06 & 2007-10-12 & \\
Latest test creation, $t_0^{c_N}$ & 2012-02-07 & 2012-04-18 & \\

\hline
\end{tabular}
\end{table}

\begin{figure*}[t]
\centering
\includegraphics[width=6in]{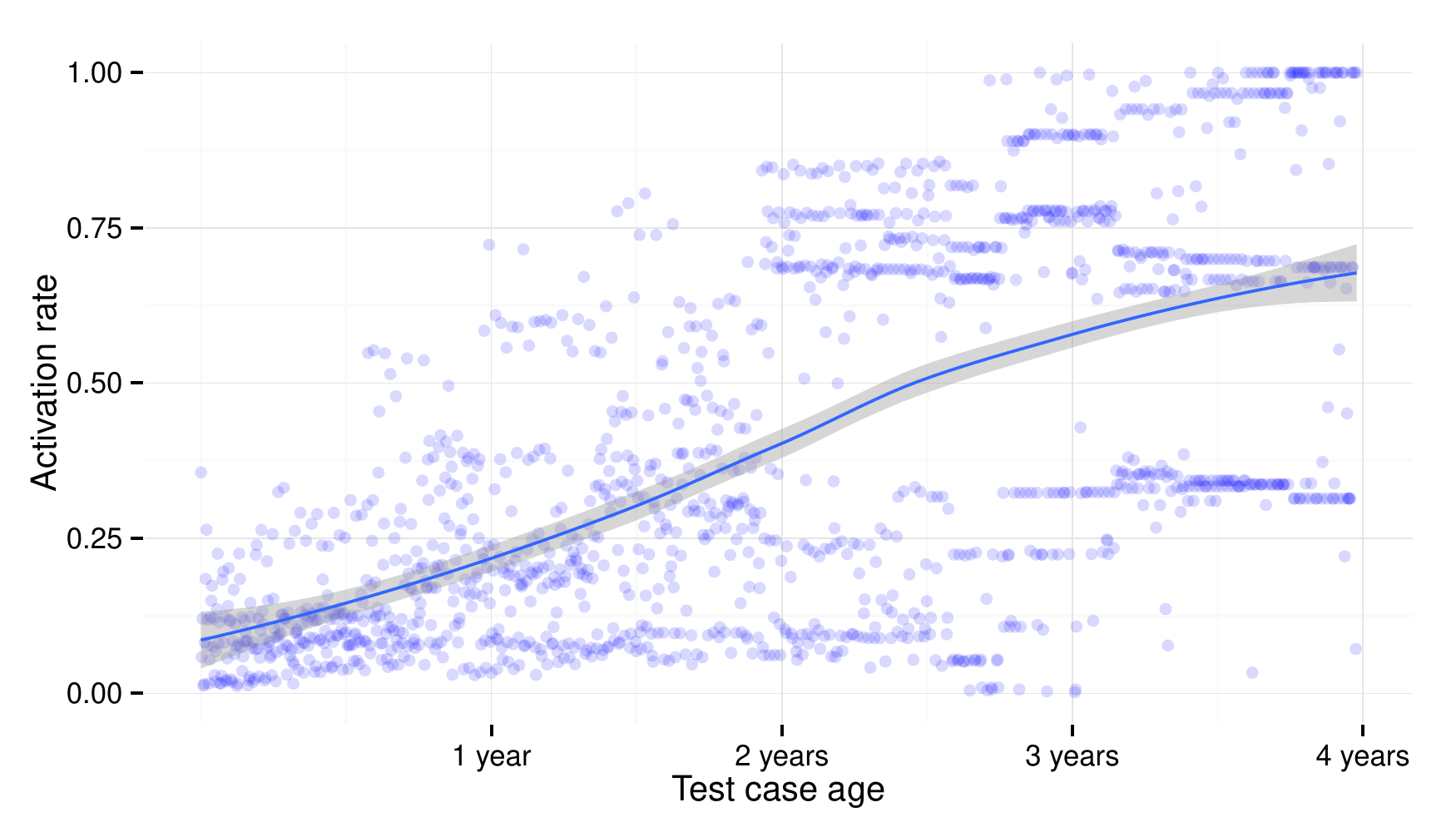}
\caption{Activation rates (blue dots) for different test case ages and a smoothed, average activation rate curve (blue with grey confidence bounds) for the \textbf{main data set} comprising 421,232 test executions of 1040 test cases.}
\label{fig_test_suite_activation_main}
\end{figure*}

The system was selected since it was one of the newest systems developed by the company and it had used more elaborate test logging procedures from early on in its development. For other systems developed by the company there was more testing data available but there was a lack of certainty about the creation times of tests. Even for the selected system we cannot be entirely sure of the actual age of test cases since the test creation time was approximated with the first registered test run in the logging database. However, test cases have clearly been added in batches over the evolution of the system which indicates either that tests are design and created in large batches and then added to the test suite or that the database logging test information lacks some test executions from early on. In discussion with test engineers at the company they consider the former explanation most likely but cannot entirely rule out the possibility that some information is missing in the logging database since it has been migrated between systems over the years.

There were a total of 5,774 test cases for the system at the time of analysis. The test cases are organized into 31 different, logical test groups that focus on different sub-systems or groups of related features. Furthermore, development of the system takes place in branches where each branch corresponds to a major release or to a specific port of the system to a different platform. Together with the main branch there is a total of nine branches. Even though the test groups are common to all branches each test execution is carried out on one specific branch, and each test case belongs to only one group.

Two data sets were extracted from the test logging database: a main data set and a validation data set. The main data set was based on the tests on the main system development branch since this has the longest continued development and thus gives the best basis for evaluating test case aging. The three largest test groups that cover distinct sets of features were selected for the main data set. The validation data set was used to ensure we do not draw wrong conclusions based on patterns that might be present by some bias in the main data set. To create the validation set we randomly selected another system branch among the 8 non-main branches and then randomly selected three test groups that had been used for this branch. Together the two selected data sets contains more than half a million test executions of a total of 1,620 unique test cases. Table~\ref{tab:datasets} presents basic information about the two data sets. Below we present our analyses of test case aging for these data sets. 

Without loss of generality, all of our analyses were carried out with the time step being days. Before applying the analyses we filtered sessions where all test cases that had been executed failed since the industrial practitioners indicated that such test sessions are likely ones in which the build failed or there was a serious problem with the testing framework itself. Only very few test sessions was of this type and we think our analysis would not have been affected even without this pre-filtering. When plotting curves and fitting models to the time series extracted from the data we also excluded days for which we had data for less than 10 test cases. Similarly, this also excluded very few data points and our analyses are likely robust enough to cope without this type of filtering altogether, at least when there is a lot of data to begin with.

\begin{figure*}[t]
\centering
\includegraphics[width=6in]{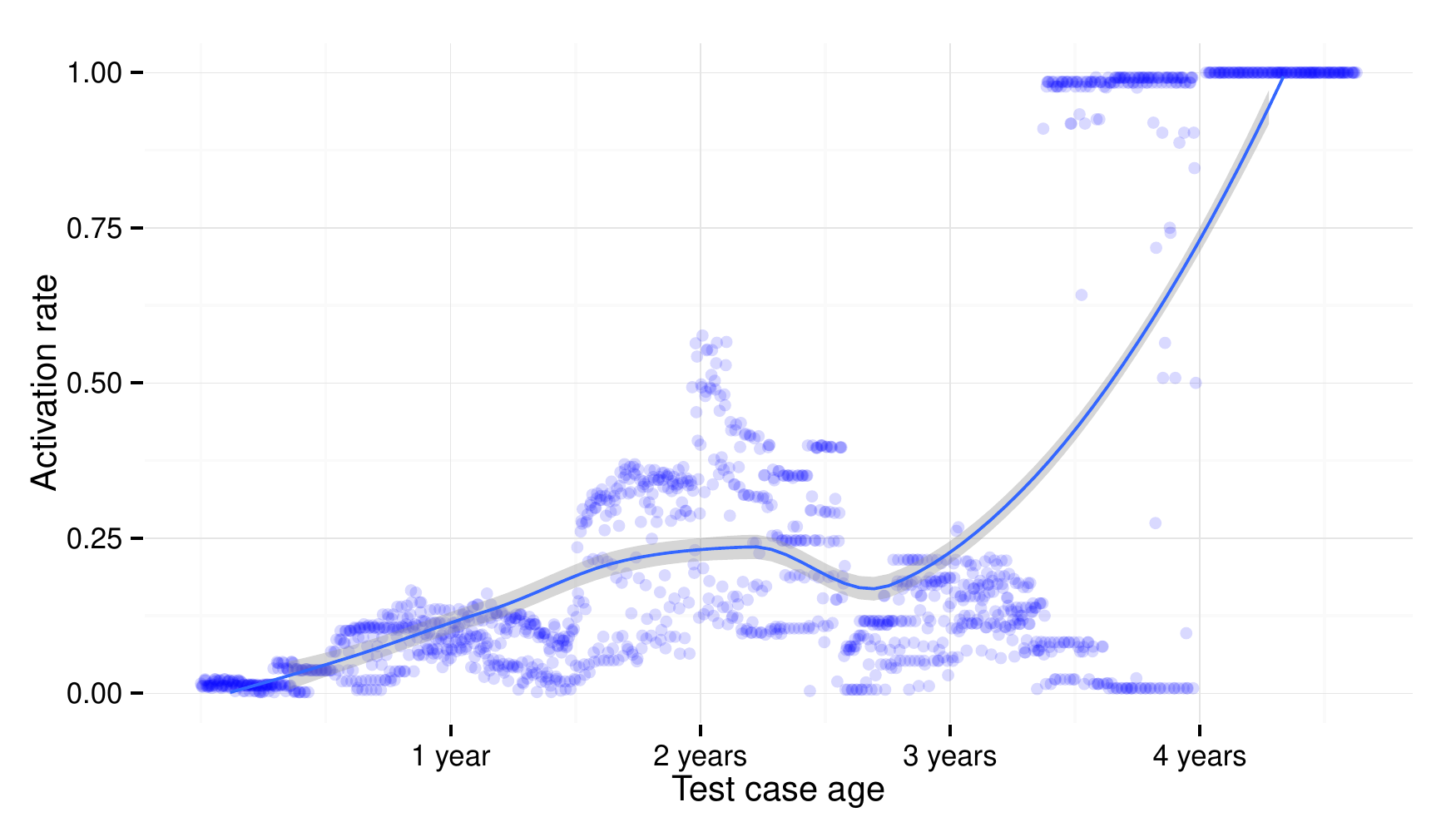}
\caption{Activation rates (blue dots) for different test case ages and a smoothed, average activation rate curve (blue with grey confidence bounds) for the (randomly selected) \textbf{validation data set} comprising 94,701 test executions of 580 test cases.}
\label{fig_test_suite_activation_validation}
\end{figure*}

\subsection{Age and aliveness of test cases}

We used two different grace periods (of 90 and 30 days, respectively) when analyzing which test cases are alive and which are dead. However, the difference in grace period had no effect on the results. For the main data set only 6.4\% of test cases, and for the validation set only 1.0\%, die, for both grace periods. Thus a large majority of test cases never die; once they have been added they continue to be used for long.

The mean age of all test cases was 1028.4 for the main and 904.3 days for the validation set, with standard deviations of 345.9 days and 366.3 days, respectively.

\framebox(228,30)[c]{%
    \parbox{220\unitlength}{Test cases that have been added to the test suite stays alive and are continuosly used.}%
}

\subsection{Activity of test cases}

Figure~\ref{fig_test_suite_activation_main} shows the empirical activation curve for the main data set and Figure~\ref{fig_test_suite_activation_validation} for the validation set. These are scatter plots of the activation rate per (age) day. 
We can see that activation levels are generally low when test cases are young but then increases over time. A possible explanation can be that young test cases have not yet been added to the nightly test runs. As test cases grow older they are more likely to be actively used.

But the patterns are not clear-cut; there seems to be a recent trend that most test cases are included, in particular for the validation data set shown in Figure~\ref{fig_test_suite_activation_validation}. Possibly the nightly test runs have been in effect only in the last year of the available interval or testing activity in general has increased over time. 

There are also bands in the activation graph. But a detailed analysis of the activation graphs require system-specific knowledge and has been left for future work. However, it is clear that in this data there is no tendency for test cases to become less actively used as they grow old.

\framebox(228,30)[c]{%
    \parbox{220\unitlength}{The activation rate of test cases does not go down over time, rather it increases.}%
}

\subsection{Overall tendency of the test case hazard curve}

\begin{figure*}[t]
\centering
\includegraphics[width=6in]{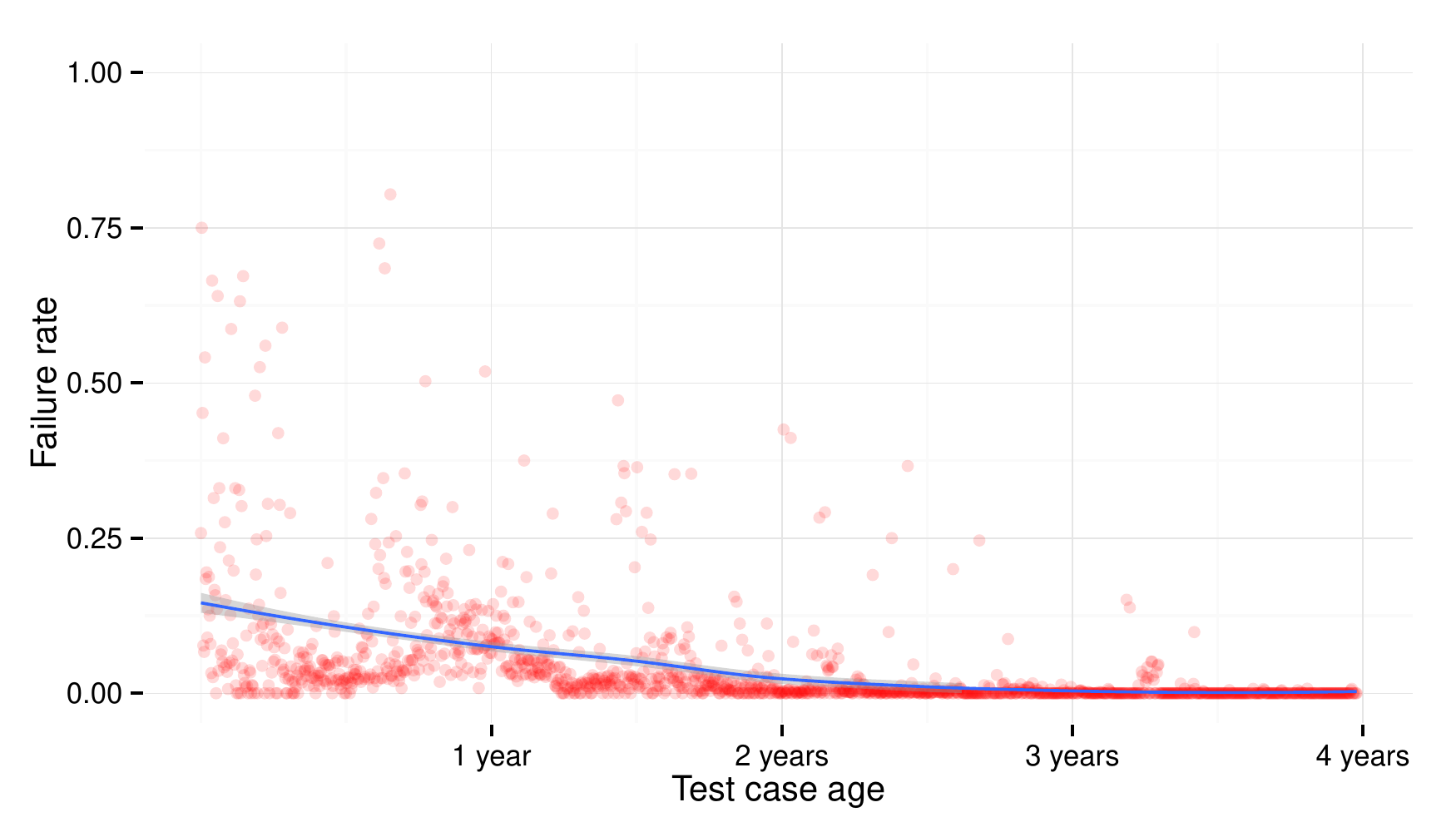}
\caption{Mean daily failure rates (red dots) for different test case ages and a smoothed, average failure rate curve (blue with grey confidence bounds) for the \textbf{main data set} comprising 421,232 test executions of 1040 test cases.}
\label{fig_test_suite_hazard_main}
\end{figure*}

The empirical test case hazard curve for the main data set, i.e. a scatter plot showing the mean daily failure rate
versus the age of a test case is shown in Figure~\ref{fig_test_suite_hazard_main}. We can see that there is quite some variation in the failure rates and for several of the days, average failure rates of more than 75\% can be seen. However, over time there is less of these high failure rates and the overall tendency is of decreasing failure rates. The overlaid blue line shows the locally smoothed (with \texttt{loess} in \texttt{R}) average curve with surrounding 95\% confidence intervals in grey.

\begin{figure*}[t]
\centering
\includegraphics[width=6in]{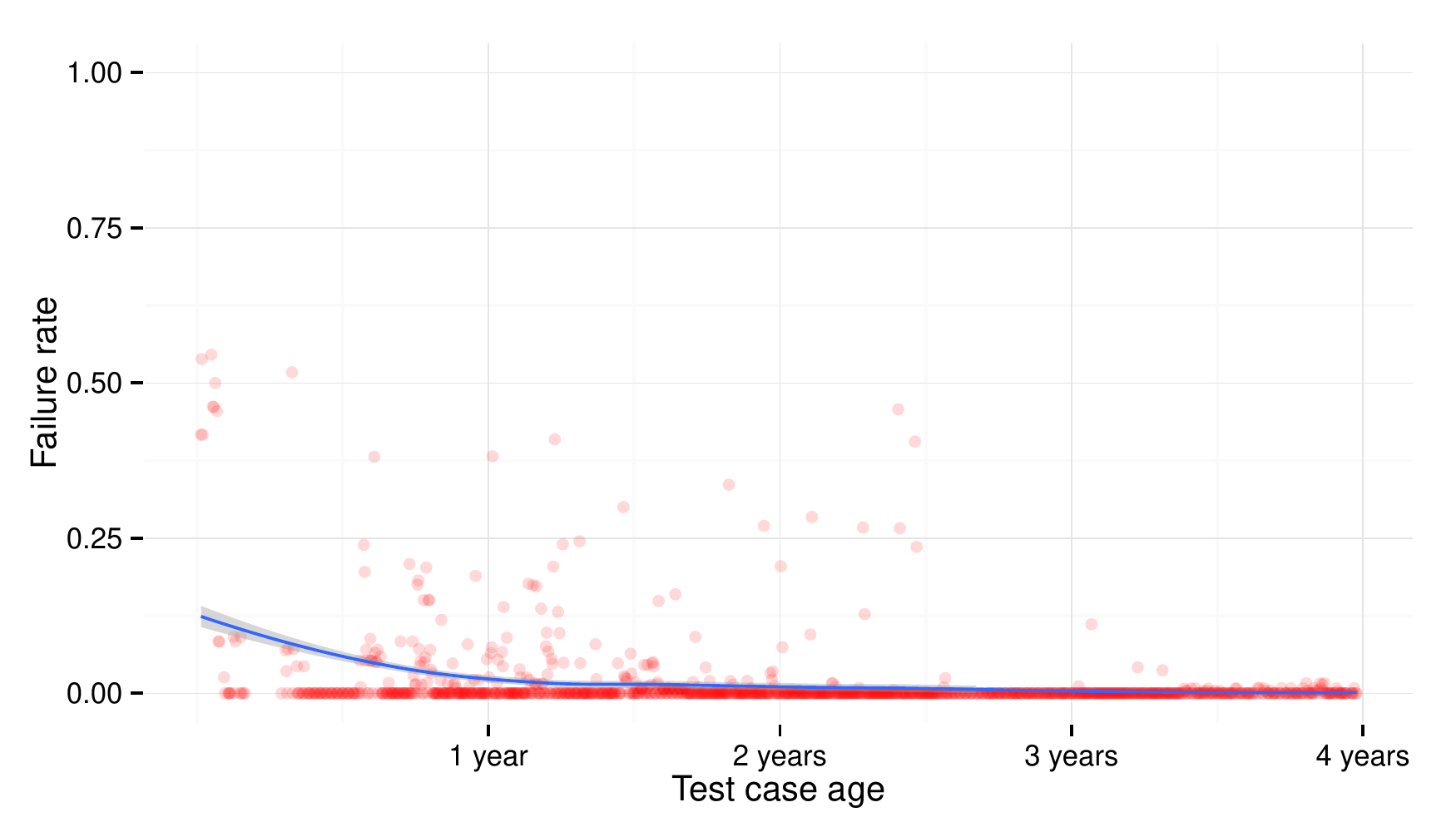}
\caption{Mean daily failure rates and a smoothed, average failure rate curve for the (randomly selected) \textbf{validation data set} comprising 94,701 test executions of 580 test cases.}
\label{fig_test_suite_hazard_validation}
\end{figure*}

The hazard curve shows a rather clear infant mortality phase with higher failure rates as the test cases are young and, correspondingly, lower failure rates and more stable behavior as test cases grow older. However, this failure rate decay phase is quite prolonged and extends all the way up to the first two to three years of testing with a test case. In actual figures the average failure rate is 11.0\% in the first year, 4.7\% in the second year, 1.3\% in the third year and only 0.3\% in the fourth year.

The high failure rates in the first and second years is probably affected by the light `bump' in failure rates that can be seen in Figure~\ref{fig_test_suite_hazard_main} around the 1 year mark. It is not clear why this type of behavior is seen in the main data set but since no such pattern is seen in the validation data set shown in Figure~\ref{fig_test_suite_hazard_validation} it is likely to be an outlier and related to one of the specific test groups. The failure rate decay is also quicker for the validation data set with an average failure rate of 4.1\% in the first year, 1.7\% in the second year, 0.8\% in the third year and 0.1\% in the fourth year.

\framebox(228,30)[c]{%
    \parbox{220\unitlength}{The failure rate of a test case is time-dependent and decreases as a test case grows older.}%
}

\subsection{Modeling the failure rate over time}

Based on the empirical hazard curves for the two data sets we can fit different statistical (regression) models to predict the failure rate given a certain test case age. We have built both linear, quadratic, cubic as well as exponential models for both data sets. The exponential model did not give any good fit and the linear model did not give a visually plausible fit. The cubic model did not improve on the quadratic. A summary of the fitted models is given in Table~\ref{tab:reg_models}. We have left out the exponential model since it gave such a poor fit and the cubic since its model takes up so much space and it does not outperform the quadratic models. All the coefficients of the models were statisticfally significant at least at a p-value of $1e^{-10}$.

\begin{table*}[!t]
\renewcommand{\arraystretch}{1.3}
\caption{Regression models for predicting the failure rate (in \%) as a function of age (in days)}
\label{tab:reg_models}
\centering
\begin{tabular}{c||c||c||c||c||c}

\hline
\bfseries Model order & \bfseries Data set & \bfseries Model & \bfseries Residual std. error & \bfseries Half-life (days) & \bfseries Half-life (months)\\

\hline\hline
Linear & main & $11.4 - 0.01*t$ &  0.08218 & 584 & 19.5\\
Quadratic & main & $14.5 - 0.02*t + 0.000009*t^2$ &  0.08095 & 373 & 12.4\\

\hline\hline
Linear & validation & $4.52 - 0.004*t$ &  0.05804 & 423 & 14.1\\
Quadratic & validation & $7.89 - 0.016*t + 0.000007*t^2$ &  0.05709 & 161 & 5.4\\

\hline
\end{tabular}
\end{table*}

Based on the predicted values of the fitted models we can estimate the half-life as the number of days of decay needed until half of the initial failure rate has vanished. We can see that for the quadratic models the half-life is over a year (373 days) for the main data set while the validation data set has a much higher decay rate with a half-life of a little over five months (161 days).


\framebox(228,40)[c]{%
    \parbox{220\unitlength}{The failure rate decline of test cases continues for the first one to two years after their creation with a half-life of between five months and a year.}%
}

\section{Validity Threats}
\label{sec:discussion}

The results of our empirical study are subject to the following threats to validity \cite{feldt2010validity}:


\textbf{Threats to \emph{external validity}} concern our ability to generalize the results of our study. In particular, since we studied only a single software system we cannot claim that the results of our empirical evaluation is generalizable. By using a large and real software system and test suite we can at least mitigate threats inherent in using small, toy software systems from academia; however, it is still only one system. Future work would need to investigate both other systems at this company but preferably also systems developed by other companies, in other countries and domains etc. Only with a large empirical base from a diverse set of companies and systems can we begin to investigate and discuss if the patterns we have identified are generally valid. The challenge is to find companies with lots of data saved and a willingness to share; this is a continuous struggle we face as researchers.

\textbf{Threats to \emph{internal validity}} concern the extent to which a causal conclusion is warranted based on our study. We have ensured the collected data has been properly extracted and the scripts used for statistical analysis has been extensively reviewed and rely on well-tested libraries for the statistical environment \texttt{R}. It is conceivable that a data set as large as the one we have studied here contains faulty data due to bugs in the testing and logging environment at the company or in how these have been used throughout the years. However, we think it is unlikely that this would have any large or adverse effects; the data itself is used by the company for their internal statistics and test process decisions. It has also been in development for many years and, most importantly, the data is extensive. We also deem any systematic biases unlikely, in particular since we have not used any specific criteria in making the selection of test group or system branch for the validation data set. Rather we randomly selected a branch and three (3) test groups.

\textbf{Threats to \emph{conclusion validity}} refers to the appropriate use of statistics to infer whether the studied variables covary. Since we have extensive data available, have used commonly used statistical procedures that are all part of a well-established and mature statistical software system, and have ensured statistical significance at strong levels this is not a big threat. It is likely that not all assumptions of the, primarily parametric, statistical tests used have been fulfilled but this is generally not a problem when there is extensive data. Even if non-parametric methods would have been used the conclusions are very likely to have been the same.


Since our study is based on the actual testing data itself and there is not really any latent variables being studied we do not consider construct validity to pose any threats to our study.

\section{Conclusions}
\label{sec:conclusions}

Software organisations often put enourmous efforts on creating high-quality test suites that they then evolve and add to. However, there is a lack of analysis methods and data on how the behavior of system test cases changes over time. In this paper we have proposed concepts and methods to perform such analyses in order to understand how test cases grow old. Collectively, we propose to call this approach `software test case aging analysis' and to, at least initially, focus on how the activation level and effectiveness of test cases change as they age. Our analysis methods have modest information requirements, that we have clearly specified in a Creation-Execution-Outcome (CEO) model, that should be available in almost all relatively mature or large software organisations. In particular, they require no access to the source code or tests themselves, only meta-data about the testing, which many organisations routinely log and have been logging for long. Our approach is based on simple statistical analysis that can be efficiently performed in any modern-day, statistical software package.

We have proposed that three aspects related to test case aging are fundamental: the age itself, and the activity and effectiveness of test cases over time. For all three concepts we have clearly defined how they are calculated from empirical data, proposed ways in which the corresponding data can be graphed and implemented the approach as scripts for the statistical envrionment \texttt{R}. Furthermore, we have proposed the concept of `test case half-life' can be used to characterise failure rate decline over time with a single number. To define the death of a test case within our restricted information model we had to introduce the concept of a grace period but found that its length did not negatively affect our analysis.

To evaluate our approach and start building an empirical knowledge base for answering questions related to test case aging we have applied our analysis methods to testing data from a large, industrial software system. We selected one large part of the testing data as the main data set and then randomly selected a smaller set for validation purposes. For the more than half million test executions of a total of more than 1500 test cases we saw that \textbf{test cases stay active but are less effective over time, i.e. \underline{they do indeed grow old} in that they are less likely to identify failures}. The decline in mean failure rate continues up to several years with a half-life of 5.4 to 12.4 months. However, they do not grow old in terms of decreased activity levels. Rather, our analyses show that very few test cases die, only between 1\% (validation data set) to 6\% (main data set), and, thus, they are active as they grow older. 

Overall, the studied test cases show a type of `infant mortality', a term commonly used to characterise the failure rates of physical components, but their infancy is quite extensive in time. However, the software test case hazard curves are not bathtub-shaped; the mean failure rate continue to fall and never raises again.

There are several ways in which our work can be extended. Foremost, we want to apply the same analyses to other, large and real-world software systems to build a more solid knowledge base which could enable the support of general statements about test case aging. In particular, we want to study if the system test cases for other systems also has similar failure rate decline rates. If there are general rules of thumb for the effectiveness of system test cases in identifying failures this can have large consequences for software testing. For one, it could provide a benchmark for how many test cases needs to be added for the test suite to stay `sharp' and `current'. But it could also be used in prioritizing regression testing efforts; old test cases are less likely to fail so should have a lower probability of being selected. 

Another way to extend our results would be to normalize the different test aging curves to the amount of testing and/or development effort that goes into the project. A third possibility would be to identify and adapt specific statistical tests to compare different periods of test case age or testing activity to each other. There is a plethora of powerful statistical methods within the time series and survival analysis fields that could be explored, for example. Lastly, more powerful test case aging analyses could be performed if the detailed evolution of the actual contents of the test cases was available in a less restrictive information model.

\section*{Acknowledgment}

The author wants to thank the many engineers and managers at the industrial company that gave me access to their systems and openly shared their development and testing experience as well as data. Without their strong support this paper would not have been possible.

\bibliographystyle{plain} 
\bibliography{paper}

\end{document}